%% file: main.tex
 \documentclass[smallabstract,smallcaptions]{dccpaper}

\usepackage{epsfig}
\usepackage{cite}
\usepackage{amsmath}
\usepackage{amssymb}
\usepackage{color}
\usepackage{url}

\newlength{\figurewidth}
\newlength{\smallfigurewidth}

\usepackage{dcolumn}% Align table columns on decimal point
\usepackage{bm}% bold math
\usepackage{comment}
\usepackage{amsmath}
\usepackage{placeins}
\usepackage{epstopdf, epsfig}
\usepackage[export]{adjustbox}
\usepackage{graphicx}
\usepackage{subfig}
\usepackage{mathrsfs}
\usepackage{booktabs}
\usepackage{multirow}
\usepackage{dcolumn}% Align table columns on decimal point
\usepackage{bm}% bold math

\begin{document}

\title
{\large
\textbf{A Physics-Informed Vector Quantized Autoencoder for Data Compression of Turbulent Flow
\thanks{This work was supported by National Science Foundation Grant No. ACI-1548562 and Office of Naval Research Grant No. N00014-18-1-2244.}}
}

% \author{%
% Mohammadreza Momenifar$^{\ast}$, Enmao Diao$^{\ast}$, Vahid Tarokh$^{\ast}$, Andrew D. Bragg$^{\ast}$\\[0.5em]
% {\small\begin{minipage}{\linewidth}\begin{center}
% \begin{tabular}{ccc}
% $^{\ast}$Institution One & \hspace*{0.5in} & $^{\dag}$Institution Two \\
% Street Address One && Street Address Two \\
% City, State, ZIP, Country && City, State, ZIP, Country\\
% \url{email@address} && \url{email@address}
% \end{tabular}
% \end{center}\end{minipage}}
% }

\author{%
Mohammadreza Momenifar$^{\ast}$$^{\dag}$, Enmao Diao$^{\dag}$, Vahid Tarokh$^{\dag}$, Andrew D. Bragg$^{\ast}$\\[0.5em]
{\small\begin{minipage}{\linewidth}\begin{center}
\begin{tabular}{c}
$^{\ast}$Department of Civil and Environmental Engineering\\
$^{\dag}$Department of Electrical and Computer Engineering\\
Duke University\\
Durham, NC, 27701, USA\\
\url{mohammadreza.momenifar@duke.edu}\\
\url{enmao.diao@duke.edu}\\
\url{vahid.tarokh@duke.edu}\\
\url{andrew.bragg@duke.edu}\\
\end{tabular}
\end{center}\end{minipage}}
}

\maketitle
\thispagestyle{empty}

\begin{abstract}
Analyzing large-scale data from simulations of turbulent flows is memory intensive, requiring significant resources. This major challenge highlights the need for data compression techniques. In this study, we apply a  physics-informed Deep Learning technique based on vector quantization to generate a discrete, low-dimensional representation of data from simulations of three-dimensional turbulent flows. The deep learning framework is composed of convolutional layers and incorporates physical constraints on the flow, such as preserving incompressibility and global statistical characteristics of the velocity gradients. The accuracy of the model is assessed using statistical, comparison-based similarity and physics-based metrics. The training data set is produced from Direct Numerical Simulation of an incompressible, statistically stationary, isotropic turbulent flow. The performance of this lossy data compression scheme is evaluated not only with unseen data from the stationary, isotropic turbulent flow, but also with data from decaying isotropic turbulence, and a Taylor-Green vortex flow. Defining the compression ratio (CR) as the ratio of original data size to the compressed one, the results show that our model based on vector quantization can offer CR $=85$ with a mean square error (MSE) of $O(10^{-3})$, and predictions that faithfully reproduce the statistics of the flow, except at the very smallest scales where there is some loss. Compared to the recent study based on a conventional autoencoder where compression is performed in a continuous space, our model improves the CR by more than $30$ percent, and reduces the MSE by an order of magnitude. Our compression model is an attractive solution for situations where fast, high quality and low-overhead encoding and decoding of large data are required. 
	
		%We perform the dimensionality reduction via extraction using Vector-Quantized Variational Autoencoder (VQ-VAE), which learns the discrete latent variables. 
% 	The performance of this lossy data compression scheme is evaluated not only with unseen data from the stationary, isotropic turbulent flow, but also with data from decaying isotropic turbulence, a Taylor-Green vortex flow, and a turbulent channel flow. Defining the compression ratio (CR) as the ratio of original data size to the compressed one, the results show that our model based on vector quantization can offer CR$=85$ with a mean square error (MSE) of $O(10^{-3})$, and predictions that faithfully reproduce the statistics of the flow, except at the very smallest scales where there is some loss. Compared to the recent study of Glaws. et. al. (Physical Review Fluids, 5(11):114602, 2020), which was based on a conventional autoencoder (where compression is performed in a continuous space), our model improves the CR by more than $30$ percent, and reduces the MSE by an order of magnitude. Our compression model is an attractive solution for situations where fast, high quality and low-overhead encoding and decoding of large data are required. 
\end{abstract}

\vspace{-0.3cm}
\section{Introduction}\label{Intro}
\vspace{-0.3cm}
\input{Intro}

\vspace{-0.3cm}
\section{Background}\label{Literature}
\vspace{-0.3cm}
\input{Literature_Background}

\vspace{-0.3cm}
\section{Method}\label{Methodology}
\vspace{-0.3cm}
\input{Methodology}

% \section{Computational Details}\label{Computational_Details}

% \input{Computational_Details}

% \section{Evaluation Metrics}\label{Evaluation_Metrics}

% \input{Evaluation_Metrics}
\vspace{-0.3cm}
\section{Experiments}\label{Results_Discussion}
\vspace{-0.3cm}
\input{Results_Discussion}
\vspace{-0.3cm}
\section{Conclusions}\label{Conclusions}
\vspace{-0.3cm}

\input{Conclusions}

% \section*{Acknowledgements}

% We would like to thank Michele Iovieno and Maurizio Carbone for kindly providing some of the DNS data used in this paper. This work used the Extreme Science and Engineering Discovery Environment (XSEDE), which is supported by National Science Foundation (NSF) under Grant No. ACI-1548562 \cite{xsede}. Specifically, the Comet cluster was used under allocation CTS170009 and the authors would like to thank Marty Kandes for his assistance with setting up the GPU environment. This work was also supported by the Office of Naval Research (ONR) under Grant No. N00014-18-1-2244.

\Section{References}
\bibliographystyle{IEEEbib}
\bibliography{References}

\end{document}

%% file: intro.tex
Turbulence is a complex dynamical system which is high-dimensional, multi-scale, non-linear, and non-local, exhibiting spatio-temporal chaotic interactions among its very wide range of scales. This has attracted substantial interest for many years both because of the intellectually stimulating challenges associated with its understanding, and also because of its practical importance to a wide range of applications such as the motion of particles in turbulent flows \cite{momenifar2018influence,momenifar2020local}, heat exchanges \cite{momenifar2015effect,nasr2015heat}, atmospheric pollution transport \cite{maxey1987gravitational}, and pumps \cite{fan2011computational, hanafizadeh2014void}. 
% Examples of such applications include those involving the motion of particles in turbulent flows \cite{momenifar2018influence,momenifar2020local} which are critical for environmental sciences such as atmospheric pollution transport \cite{maxey1987gravitational}, cloud formation \cite{beard1993warm}, volcanic eruptions \cite{ongaro2007parallel}, orographic rainfall \cite{eghdami2019extreme} and combustion processes \cite{kuo2012fundamentals}, and the effect of turbulence on the design and performance of engineering devices such as wind turbines \cite{abdulqadir2017physical, moriarty2002effect}, airplane design \cite{etkin1981turbulent}, heat exchangers \cite{bianco2011numerical,momenifar2015effect,nasr2015heat}, and pumps \cite{fan2011computational, hanafizadeh2014void}, to name just a few.  

The growing availability of computational resources in recent decades has facilitated the exploration and understanding of turbulent flows. From the Computational Fluid Dynamics (CFD) perspective, high-fidelity simulations of turbulent flows may be achieved by solving the Navier-Stokes (NS) equation numerically, a procedure known as Direct Numerical Simulation (DNS). A less expensive computational approach is to use Large Eddy simulation (LES) in which the small scales are modeled rather than simulated, reducing computational cost at the expense of accuracy and access to small-scale flow information. Both DNS and LES require significant computational resources, and can produce large amounts of data that is cumbersome to store, transfer (bandwidth requirements) and analyze. Such challenges can be addressed by the development of efficient and accurate data compression techniques.   
% \par \smallskip

Data compression in this CFD context may be employed for both compressed data checkpointing (storing the compressed data and using it for restarting the simulation), and post-processing purposes.   
The main goal of such compression processes is to truncate the size of the data files while also ensuring that (i) simulation restarts using the compressed data would not significantly impact the long-term behavior of the simulated flow (ii) the compressed data still preserves the essential statistical properties of the turbulent flow with reasonable accuracy. 
% Compression techniques can be broadly classified into lossless and lossy, depending on the information loss during a data reduction process. In dealing with large CFD simulations, often a lossy compression algorithm is favored, due to its reduced memory requirements while offering higher compression capabilities, although this approach sacrifices the ability to retrieve the original data. Recent studies in the image compression community \cite{diao2020drasic,johnston2018improved} have shown that machine-learning based lossy compression techniques, particularly those based on deep neural networks, can outperform classical methods such as JPEG standard \cite{wallace1992jpeg} and BPG \cite{bellard2015bpg}. 
% \par \smallskip

Machine learning techniques are bringing fresh perspectives in many areas. Among machine learning techniques, deep learning models have received significant attention due to their ability to capture complex interactions and achieve outstanding performance across a wide range of applications in information technology, healthcare and engineering, to name a few. In the context of data compression, common deep neural networks are based on an autoencoder architecture, which has also gained interest in the CFD community for the purpose of data compression of three dimensional turbulent flows \cite{glaws2020deep}, \cite{mohan2020spatio}.
Our study is in line with these recent endeavors and proposes a deep learning based data compression framework that compresses/encodes the velocity field data of three dimensional turbulent flows into a discrete space through quantization of the bottleneck representation, the so-called vector-quantized autoencoder. This approach involves a calibration process that enables infusing prior knowledge of the data to boost the performance of model. Compared to the recent autoencoder network proposed in \cite{glaws2020deep}, training our framework is computationally much cheaper and the trained model significantly improves not only the compression ratio but also the accuracy of the reconstructed data. 

%To the best of our knowledge, this is the first deep-learning framework based on the concept of qunatization in the turbulence literature and its performance is evaluated with rigorous turbulence-based metrics on different turbulent flows.    
%\par \smallskip

% This outline of this paper is given next. A  brief literature review on the data compression techniques, application of deep learning in turbulence research and data compression is provided in Section \ref{Literature}. In Sections \ref{Methodology} our mathematical methodology, and in  Section \ref{Computational_Details} our implementation approach is discussed.  Assessment metrics and methodologies are described in Section \ref{Evaluation_Metrics}. Using these metrics, numerical results are provided  in Section \ref{Results_Discussion} evaluating the performance of our model on several test cases. These test cases represent turbulent flows with different characteristics from the training set data. Finally, Section \ref{Conclusions} provides a brief summary of our findings.

%% file: Literature_Background.tex
The analysis of large turbulent flow simulations is generally more concentrated on the statistical quantities of flow field rather than its instantaneous values. Therefore, we prefer a lossy compression scheme that can extract the most relevant physics of turbulence hence resulting in a minimal impact on the quality of statistics of post-processed data from a physics point of view. Given that loss of information is unavoidable in lossy compression methods, we desire a model that can offer a controllable level of trade-off between the compression ratio and distortion of data, independent of input data. Lossy compression schemes involve two steps of (i) compression, in which original data is encoded into a latent space and (ii) decompression, in which the compressed data is decoded. The quality of these lossy dimension reduction models is evaluated based on their compression capability, which is measured by computing compression ratio (CR) and defined as:
\begin{align}\label{eq:CR}
% CR = \frac{original \: file \: size}{compressed \: file \: size}
\text{CR} = \frac{\text{original file size}}{\text{compressed file size}}
\end{align}
and its performance in reconstructing the original data, which is measured by different metrics depending on the applications. 
With the rise of deep learning and the rapid development of its theory, its applications in the field of image compression have proven remarkably successful. Common deep neural networks for the purpose of image compression have an auto-encoder architecture, including recurrent and non-recurrent auto-encoders, and are mainly composed of CNN layers \cite{diao2020drasic}. The auto-encoder architecture is composed of two networks, namely the encoder and decoder. The encoder performs down-sampling on the input data to generate a compressed non-linear latent space (also called the bottleneck) representation of the input, while the decoder takes the output of the encoder and reconstructs the input data via upsampling. 

Convolutional Neural Network-based autoencoder has also received attention in the turbulence community for use in dimension reduction of velocity field data from three-dimensional turbulent flows. In the recent study of \cite{mohan2020spatio}, a convolutional autoencoder was employed to obtain a low dimensional representation of a three-dimensional velocity field from an isotropic turbulent flow. The main purpose of their study was to model spatio-temporal realizations of isotropic turbulence. They constructed a framework that offers a compression ratio of $125$. Their compression results indicate that their model can capture large scales of flow accurately and inertial scales with some distortion, but it failed drastically in preserving the small scales of flow, as seen in their compression model results for the turbulence energy spectrum and the probability distributions of the longitudinal velocity gradients. The recent turbulence data compression study of \cite{glaws2020deep} centered around the outperformance of their CNN based autoencoder against a variant of singular value decomposition (SVD) for the purpose of in-situ data compression, with a focus on generating lossy restart data. Their autoencoder offers a compression ratio of $64$ and has been trained on decaying isotropic turbulence, and then tested on Taylor-Green vortex and pressure-driven channel flows. Their findings clearly demonstrate the remarkable performance of their model in reconstructing physical characteristics of flows which were not seen by the model during training. Our study is motivated by these recent works to design a data compression model that increase not only the compression ratio but also the performance of the reconstructed data. 

% Their results for the lossy restart data show that the trained network produces compressed data files that, when used as restart files in simulations, leads to simulation results that preserve the important flow properties when compared with the results generated using the original, full data for the restart files. Nevertheless, although their model does a much better job at capturing the properties of the inertial range scales compared to the model in \cite{mohan2020spatio}, it still it leads to a noticeable loss of information for the small scales of flow. 
% The results of these recent studies indicate that further efforts in using CNN-based autoencoders for data compression of turbulent flows can be very fruitful. 

% In section \ref{Methodology}, we elaborate on the formulation underlying the model, and in section \ref{Computational_Details} we discuss the details of the implementation and computations. 

%% file: Methodology.tex
The main difference between our framework and previous studies is that we incorporate the concept of vector quantization and generate a bottleneck representation in a discrete space (an integer representation). Furthermore, we infuse prior physical knowledge of the data into the model to enhance the ability of model in respecting the small-scale characteristics of the turbulence which were not well-captured in previous works. Finally, additional (off-line) training of a spatio-temporal model over such a discrete latent space is $\text{CR} \times $ faster, allowing the model to be trained efficiently on higher resolution data. To the best of our knowledge, this is the first deep learning framework embedded with the quantization concept in the turbulence literature. Our results indicate that our framework outperforms the existing models while not suffering from their shortcomings of capturing small scales.

\subsection{Vector-Quantized Autoencoder}
A Vector-Quantized (fully convolutional) autoencoder encodes the input data in a discrete latent space and can effectively use the capacity of latent space by conserving important features of data that usually span many dimensions in data space (such as objects in images) and reducing entropy (putting less focus on noise) \cite{van2017neural}. Given that turbulence is inherently non-local (its features span many dimensions in the computational domain), VQ might be a good tool to retain import characteristics of turbulence during compression of the data. 
Compared to a conventional autoencoder, a Vector-Quantized Autoencoder has an additional Vector-Quantizer module. The encoder ($E$) serves as a non-linear function that maps input data ($x$) to a vector $E(x)$. The quantizer module takes this vector and outputs an index ($k$) corresponding to the closest codeword in the codebook to this vector ($e_{k}$):
\begin{align}\label{eq:Quantizer}
\text{Quantize}(E(x)) = e_{k},\, \text{where} \; k = \underset{j}{\arg\min} \parallel E(x) - e_{j} \parallel_{2}. 
\end{align}
Codeword index $k$ is used for the integer representation of the latent space, and $e_{k}$ serves as the input of decoder ($D$) which operates as another non-linear function to reconstruct the input data. The Vector-Quantizer module brings two additional terms in the loss function, namely codebook loss and commitment loss, to align the encoder output with the vector space of the codebook. The entire VQ-AE loss is defined as:
\begin{align}\label{eq:VQ_VAE_Loss}
\mathcal{L}(x,D(e))=\underbrace{\parallel x - D(e) \parallel^{2}_{2}}_{reconstruction~loss} + \underbrace{\parallel sg\{E(x)\} - e \parallel^{2}_{2}}_{codebook~loss} + \underbrace{\beta \parallel sg\{e\} - E(x) \parallel^{2}_{2}}_{commitment~loss}. 
\end{align}
In the above, the reconstruction loss trains both encoder and decoder parameters where 
the gradient of reconstruction error is back-propagated to the decoder first, and then directly passed to the output of the encoder using the straight-through gradient estimator \cite{bengio2013estimating} (because there is no real gradient for the \emph{$\arg\min$} term in Equation (\ref{eq:Quantizer})). The codebook loss trains only the codebook by moving the picked codeword $e_{k}$ towards the output of the encoder. The commitment loss trains only the encoder by encouraging the output of the encoder to be close to the selected codeword, and to avoid jumping frequently between different codewords in the codebook. The $\beta$ coefficient represents the commitment coefficient which controls the reluctance against this fluctuation, and $sg\{\cdot\}$ denotes the stop gradient operator which does not propagate gradients with respect to its arguments. For the codebook design, the codebook loss can be replaced with the exponential moving average scheme to update codebook parameters. More details can be found in \cite{van2017neural,roy2018theory, razavi2019generating}.

\subsection{Incorporating prior knowledge of data into framework}\label{Incorporating_prior_knowledge}
So far we have highlighted two connections between the characteristics of turbulence and the proposed framework: (i) the compositional hierarchy of convolutional networks and multiscale nature of turbulence. (ii) the capability of VQ to capture spatially correlated features of the flow that also exist between different scales. These are general characteristics of many multi-scale physical phenomenon. Consequently, the above framework can be employed for other applications such as climatology, geophysics, oceanography, astronomy, and astrophysics where the problem of interest is understanding patterns and correlations. We employ this framework for the case of turbulent flows. It would be beneficial if we could infuse our prior knowledge of input data into the model so that it is enforced to obey those constraints. To this end, we impose such constraints by additional regularization terms in the loss function of the model. 

As noted earlier, preserving small-scale properties of the turbulent flow was a challenge for prior compression models.  It may be of interest to add appropriate constraints in order to capture these more faithfully. Given that our model will be trained on isotropic turbulence, the appropriate constraints for this kind of flow will be our main focus here. Let us consider the Cartesian components of the velocity gradient tensor, $A_{ij}=\partial u_i/\partial x_j$. The incompressibility of the flow implies that $A_{ii} = 0$. Furthermore, ``Betchov relations'' \cite{betchov1956inequality} for an incompressible, statistically homogeneous turbulent flow are given by
\begin{align}\label{eq:Betchov_1}
\langle S_{ij} S_{ij} \rangle &= \langle R_{ij} R_{ij} \rangle = \frac{1}{2}\langle \omega_{i} \omega_{i} \rangle \\
\label{eq:Betchov_2}
\langle S_{ik} S_{kj} S_{ij}\rangle &= -\frac{3}{4}\langle S_{ij} \omega_{i} \omega_{j} \rangle
\end{align}
where $S_{ij} \equiv (1/2)(A_{ij}+A_{ji})$ is the strain-rate, $R_{ij} \equiv (1/2)(A_{ij}-A_{ji})$ is the rotation-rate, and $\omega_{i} = \epsilon_{ijk}R_{jk}$is the vorticity (where $\epsilon_{ijk}$ is the Levi-Civita symbol). We summarize all these constrains as:
\begin{align}\label{eq:VG_Constraint}
&\text{Velocity Gradient Constraint (VGC)} = \underbrace{\text{MSE}(A_{ij},\widehat{A_{ij}})}_{i= j} + a \times \underbrace{\text{MSE}(A_{ij},\widehat{A_{ij}})}_{i\neq j}\\
&\text{Higher Order Constraints (HOC)} =  \text{MAE}(\langle S_{ij} S_{ij} \rangle,\widehat{\langle S_{ij} S_{ij} \rangle}) + 
\text{MAE}(\langle R_{ij} R_{ij} \rangle,\widehat{\langle R_{ij} R_{ij} \rangle}) + \nonumber\\
&\text{MAE}(\langle S_{ik} S_{kj} S_{ij} \rangle,\widehat{\langle S_{ik} S_{kj} S_{ij} \rangle})+
\text{MAE}(\langle S_{ij} \omega_{i} \omega_{j} \rangle,\widehat{\langle S_{ij} \omega_{i} \omega_{j} \rangle}),
\label{eq:Other_constraints}
\end{align}
\noindent where for each quantity of interest, $A_{ij}$,
$\langle S_{ij} S_{ij} \rangle$, $\langle R_{ij} R_{ij} \rangle$, $\langle S_{ik} S_{kj} S_{ij} \rangle$, and $\langle S_{ij} \omega_{i} \omega_{j} \rangle$
the reconstructed ones are denoted by $\widehat{A_{ij}}$, $\widehat{\langle S_{ij} S_{ij} \rangle}$, $\widehat{\langle R_{ij} R_{ij} \rangle}$, $\widehat{\langle S_{ik} S_{kj} S_{ij} \rangle}$, and $\widehat{\langle S_{ij} \omega_{i} \omega_{j} \rangle}$, respectively. The coefficient $a$ is introduced in equation \ref{eq:VG_Constraint} to account for the differences in the statistics of the longitudinal and transverse components. For example, at the small-scales of isotropic turbulence, the variance of the transverse velocity gradients are twice the size of the longitudinal ones \cite{pope_2000}. We penalize the deviations in the Equation (\ref{eq:Other_constraints}) with mean absolute error (MAE), which is less sever than the MSE in the Equation (\ref{eq:VG_Constraint}), in order to put less focus on this term as it is a secondary objective (usually the error at high-order statistics are large and we mainly want to focus on recovering the velocity field and its first-order statistics).
Finally adding these constraints as regularization terms to VQ-AE loss function gives the overall loss function (OL) given below
\begin{align}\label{eq:Final_loss}
\text{Overall Loss (OL) } = \text{VQ-AE loss} + \alpha \times \text{VGC} + \gamma \times \text{HOC}.
\end{align}
% which will be use for training.
%

%% file: Results_Discussion.tex
\subsection{Experimental Setup}
\begin{figure}
	\centering
	\vspace{-0.3cm}
	\includegraphics[width=0.7\linewidth]{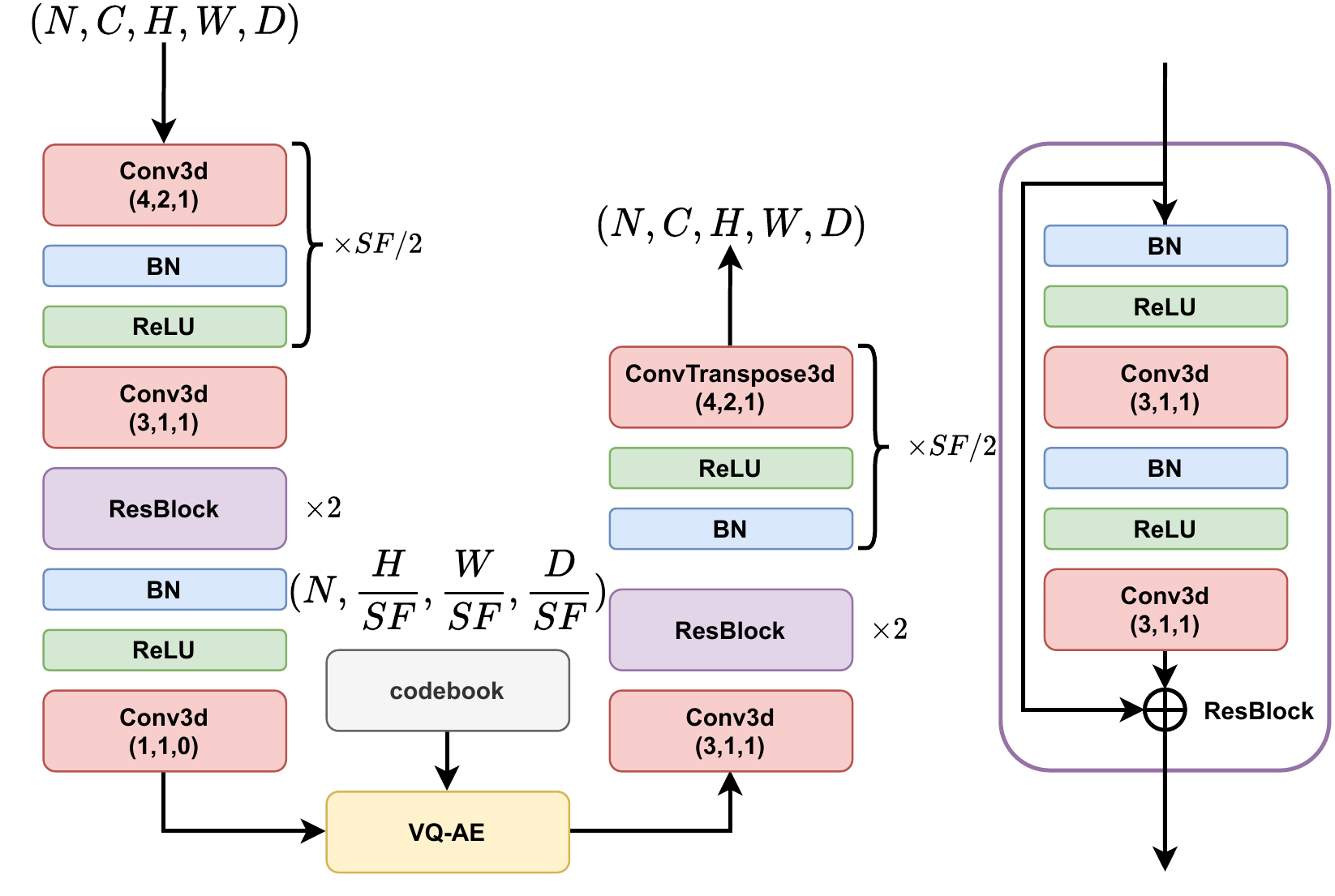}
	\vspace{-0.3cm}
	\caption{Schematic of the VQ-AE architecture. In this figure, $N$ is the batch size, $C$ is the input channel size, $H$, $W$ and $D$ represent the height, width and depth of input data, and $SF \in \{2,4,8\}$ is the scaling factor. BN represnets Batch Normalization and  Conv3d (4,2,1) represents a 3d Convolution layer with a kernel of size $4^3$, stride $2$ and padding $1$.}
	\label{fig:VQ-AE_Schematic}
	\vspace{-0.3cm}
\end{figure}

\textbf{Data} \; We train our model using high-fidelity DNS data of a three-dimensional, statistically stationary, isotropic turbulent flow with Taylor-Reynolds number $R_\lambda = 90$, solved on a cubic domain with $128$ grid points in each direction. More details on the DNS can be found in our previous works \cite{ireland2013highly,momenifar2020local,momenifar2018influence}. In our training data set we have $40$ snapshots (as opposed to $3000$ in the recent study of \cite{glaws2020deep}) equally spaced in time, covering a time span of $2T_{l}$ ($T_{l}$ denotes large eddy turn over time). Each snapshot consists of the three components of the velocity field ($u , v, w$) at all $128^3$ grid points. We test the performance of the trained framework on a variety of flows to determine how well the model can compress flows different from those used in the training. For comparison purposes, the flows are selected to correspond to those considered in \cite{glaws2020deep}. Specifically, we start with statistically stationary isotropic turbulence, but then consider decaying isotropic turbulence, and a Taylor-Green vortex flow.\\
\textbf{Model} \; The schematic of this framework is shown in Figure~\ref{fig:VQ-AE_Schematic}. We design our network so that it can transform and downsample original data by a scaling factor of $SF \in \{2,4,8\}$ depending on the level of reconstruction quality and compression needed. Indeed, an input data of shape $(3,128,128,128)$ is compressed to $(1,64,64,64)$ with $SF=2$, $(1,32,32,32)$ with $SF=4$, or $(1,16,16,16)$ with $SF=8$. With $K = 512$ representing the size of the codebook (number of codewords) and mapping three velocity components into one in the discrete latent space, we can achieve $\frac{3 \times 32}{1 \times 9} \times (SF)^{3}$ reduction in bits, corresponding to $\text{CR} =\{85,683,5461\}$ respectively. We set the weights of regularization $\alpha=0.1$ and $\gamma=10^{-3}$. 
% The test snapshot of stationary isotropic turbulence comes from the same DNS that was used for training the model, but this snapshot is from a time that is $2T_{l}$ beyond the time of the last snapshot used in the training set. We obtained the snapshot for decaying isotropic turbulence from \cite{glaws2020deep}, which represents a flow with $R_\lambda = 89$ simulated on a $128^3$ grid. Our Taylor-Green vortex snapshot comes from a DNS with $Re = 1600$ (where $\nu = 1/1600$) performed on a $192^3$ grid, similar to that used in  \cite{glaws2020deep}.

\begin{table}[hbtp]
\centering
\caption{Summary of the performance on unseen data from statistically stationary isotropic, decaying  isotropic, and decaying Taylor-Green vortex turbulence.}
\vspace{-0.2cm}
\label{tab:comparison}
\resizebox{0.7\columnwidth}{!}{
\begin{tabular}{@{}ccccccc@{}}
\toprule
Turbulent Flow & CR   & Method & MSE             & MAE             & MSSIM          & HOC            \\ \midrule
\multirow{3}{*}{\begin{tabular}[c]{@{}c@{}}stationary\\ isotropic\end{tabular}} &
  85 &
  \multirow{3}{*}{VQ-AE} &
  \textbf{0.0044} &
  \textbf{0.0499} &
  \textbf{0.977} &
  \textbf{18.37} \\
               & 683  &        & 0.0201          & 0.1070          & 0.909          & 51.25         \\
               & 5461 &        & 0.1900          & 0.3240          & 0.600          & 112.59        \\ \midrule
\multirow{5}{*}{\begin{tabular}[c]{@{}c@{}}decaying\\ isotropic\end{tabular}} &
  64 &
  SVD\cite{glaws2020deep} &
  2.8043 &
  2.2944 &
  0.198 &
  \multirow{2}{*}{N/A} \\
               & 64   & AE\cite{glaws2020deep}     & 0.0865          & 0.3744          & 0.946          &               \\
 &
  85 &
  \multirow{3}{*}{VQ-AE} &
  \textbf{0.0018} &
  \textbf{0.0326} &
  \textbf{0.970} &
  \textbf{9.81} \\
               & 683  &        & 0.0080          & 0.0693          & 0.882          & 20.39         \\
               & 5461 &        & 0.0504          & 0.1720          & 0.598          & 37.83         \\ \midrule
\multirow{3}{*}{\begin{tabular}[c]{@{}c@{}}decaying\\ Taylor-Green\\ vortex\end{tabular}} &
  64 &
  SVD\cite{glaws2020deep} &
  0.0253 &
  0.2112 &
  0.398 &
  \multirow{2}{*}{N/A} \\
               & 64   & AE\cite{glaws2020deep}     & \textbf{0.0017} & 0.0483          & \textbf{0.953} &               \\
               & 85   & VQ-AE  & 0.0027          & \textbf{0.0395} & 0.830          & \textbf{1.33} \\ \bottomrule
\end{tabular}
}
\end{table}
\subsection{Experimental Results}
\textbf{Comparison with the state-of-the art} \; In Table~\ref{tab:comparison}, we summarize our quantitative results with respect to the performance of the trained VQ-AE model on an unseen realization from statistically stationary isotropic, decaying isotropic, and Taylor-Green vortex turbulence flow. We also compare our results with the state-of-the-art \cite{glaws2020deep} using SVD and AE. The results of $SF=2, \text{CR}=85$ indicate that our model can capture up to third order statistics of the velocity components with reasonable accuracy while offering $85 \times$ compression, meaning an $85 \times $ increase in data transfer and $85 \times $ decrease in disk usage. ML-based compression methods lead to loss of information mainly at the smallest scales was also observed in \cite{glaws2020deep,mohan2020spatio}. Our model with $SF=4, \text{CR}=683$ captures most of the information content of the flow and may be employed for the situations where there is less interest in accurate representations of the smallest scales of the flow, i.e. flows with relatively little kinetic energy. Similarly, our model with $SF=8, \text{CR}=5461$ may be used when the main interest is in accurately representing the large scales of flow. 

In the study of \cite{glaws2020deep}, they used decaying isotropic turbulence flow as training data and proposed to use fully convolutional AE model with $\text{CR}=64$. On the contrary, we use stationary isotropic turbulence flow as training data and test on decaying isotropic turbulence flow. Compared with the results of decaying isotropic turbulence flow presented in \cite{glaws2020deep}, our $SF=2$ model improves both the MSE and MAE by an order of magnitude, and the MSSIM by 2\%, while offering a $\text{CR}=85$ which corresponds to more than a 30\% enhancement in the compressive capabilities. Our results also demonstrate that loss of information occurs at smallest scales of flow, indicated by Higher-Order Constraints (HOC), and is enhanced as we increase CR. As for the results of decaying Taylor-Green vortex turbulence flow, although we could not access the exact flow snapshot used in the study of \cite{glaws2020deep}, to make a fair comparison we test our model on multiple snapshots to verify that our model has a robust performance across different realizations. Our approach performs competitively with the baseline method with respect to the point-wise metrics (MSE and MAE) but 10 \% lower MSSIM. It may due to that the baseline method randomly chooses the values of the Taylor-scale Reynolds to simulate augmented training data and thus enhance the generalization ability of model \cite{glaws2020deep}. However, our method can reconstruct more accurate physics-based statistics for decaying Taylor-Green vortex turbulence flow compared with other types of flow. 

\begin{figure}[hbtp]
	\centering
	\vspace{-0.3cm}
	\includegraphics[width=0.8\linewidth]{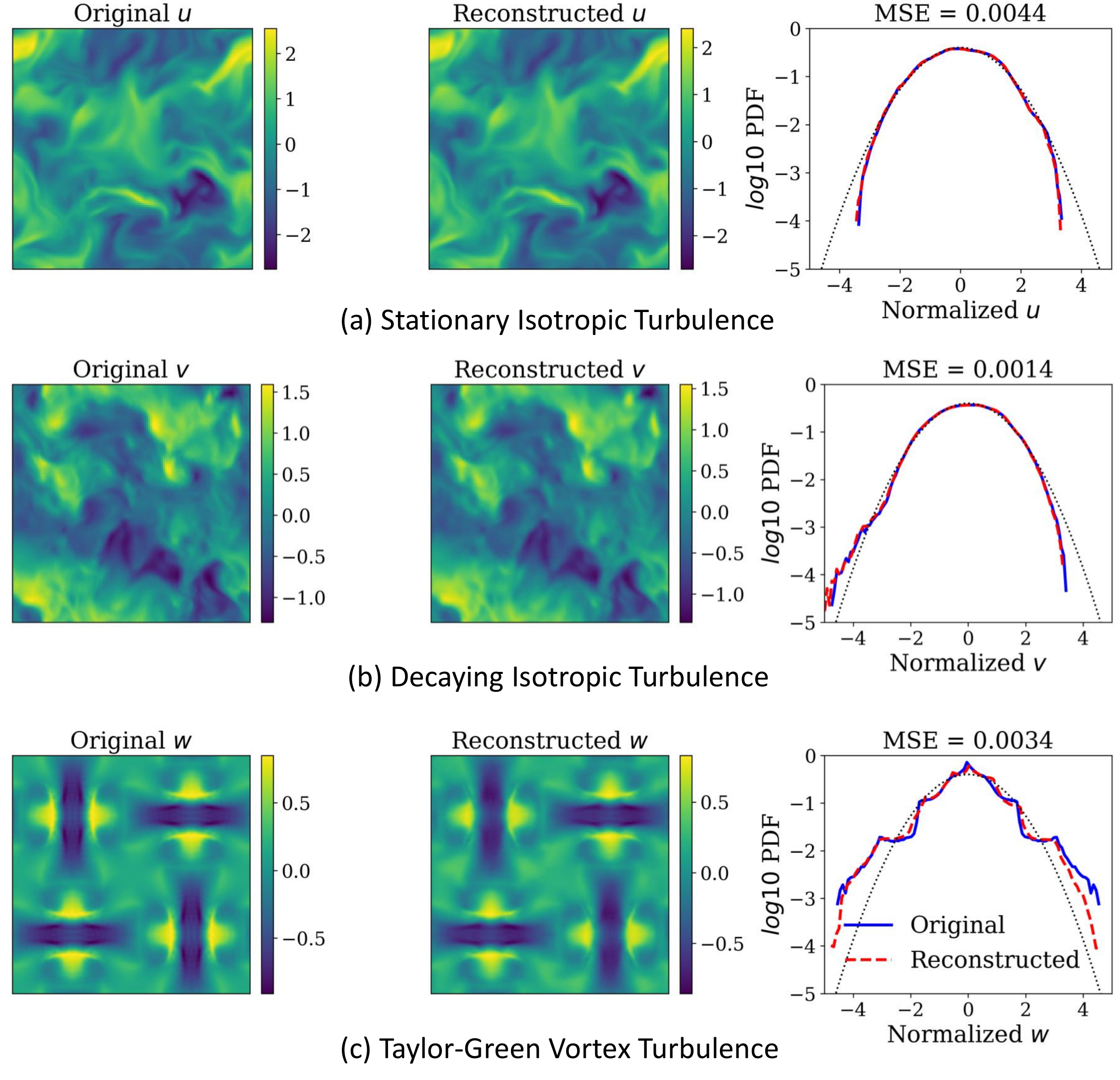}
	\vspace{-0.3cm}
	\caption{Comparing original and reconstructed 3D (a) stationary isotropic, (b) decaying isotropic, and (c) Taylor-Green vortex turbulence compressed by VQ-AE with $SF=2, \text{CR}=85$, 2D snapshots, and PDFs of velocity components $u,v,w$.}
	\label{fig:velocity}
	\vspace{-0.3cm}
\end{figure}

\begin{figure}[hbtp]
	\centering
	\vspace{-0.3cm}
	\includegraphics[width=0.8\linewidth]{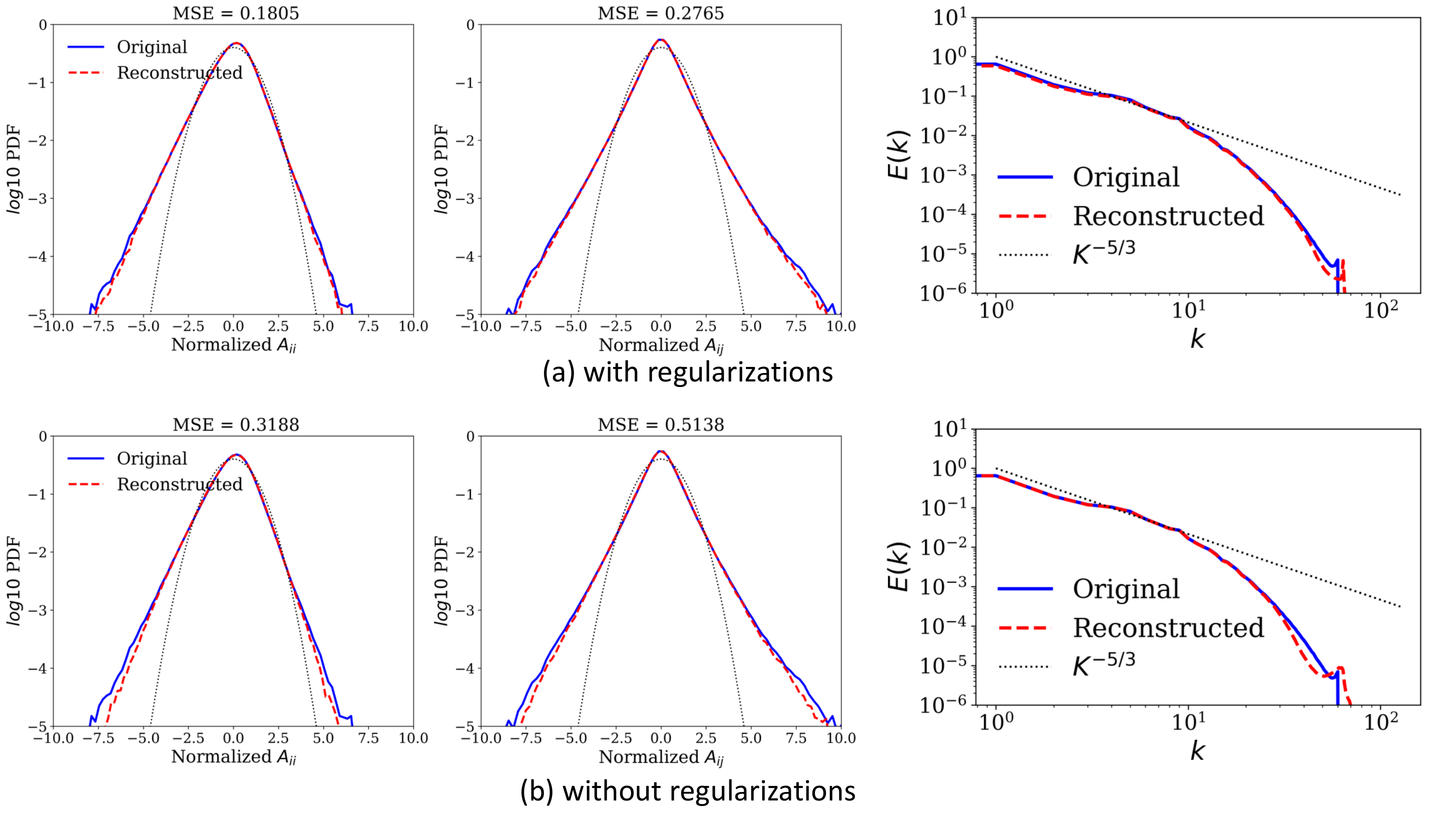}
	\vspace{-0.3cm}
	\caption{(a) with and (b) without regularizations for PDFs of normalized longitudinal (left), transverse (middle) components of velocity gradient tensor $\boldsymbol{A}$, and Turbulence Kinetic Energy spectra (right) of stationary isotropic turbulence flow.}
	\label{fig:regularization}
	\vspace{-0.3cm}
\end{figure}

We illustrate the qualitative results of velocity components and its probability density functions (PDFs) in Figure~\ref{fig:velocity}. We show original and reconstructed two-dimensional snapshots of three velocity gradient components $u, v, w$ from stationary isotropic, decaying isotropic, and Taylor-Green vortex turbulence flow respectively. The snapshot comparisons show that our model captures very well the instantaneous spatial structure of the flow. The PDF results show that the model accurately captures the statistical properties of the velocity field. Compared to the results in the study of \cite{mohan2020spatio} and \cite{glaws2020deep}, our model more accurately captures the tails of the PDFs with minor deviations in the far tails. \\
\textbf{Effect of Regularizations}\; To better understand the effect of calibrating the loss function by incorporating prior knowledge of the properties of the data, we trained another model without those regularization terms, which amounts to setting $\alpha = \gamma =0$ in the model. As shown in Figure~\ref{fig:regularization},
including the regularization term enhances the ability of the model to capture the intermittent fluctuations of the velocity gradient characterized by the tails of the PDFs. Furthermore, the results of the Turbulence Kinetic Energy spectra demonstrate that including the regularization term makes some improvement at the smallest scales (highest wavenumbers). The deviation at the smallest wavenumbers (large scales) can be attributed to the fact that in the original data, the energy content at these scales is very small, $O(10^{-32})$, and so it is hard for the model to retain such precision in the reconstructed data. Nevertheless, the model without regularization already performs well at capturing the properties of the turbulent flow.

%% file: Conclusions.tex
In this study, we have proposed a vector quantized deep learning framework, the so called vector quantized autoencoder or VQ-AE, for the compression of data from turbulent flow simulations. We have calibrated the loss function of the model to infuse prior knowledge of the flow in the form of constraints in order to boost the model performance. Our data compression model can facilitate the dynamical modeling of three dimensional turbulence as it provides a latent space with high information content that can be used in a sequence modeling deep learning framework to learn the spatio-temporal characteristics of flow. Furthermore, we believe this data compression framework is not limited to Computational
Fluid Dynamics (CFD) simulations but can be easily applied to compress data from other complex physical simulations with structured mesh domains. 